\documentclass{article}
\usepackage{amsmath,latexsym,exscale,amssymb}
\input epsf
\newcommand {\z }   {\zeta}

          \newcommand {\x }  {\xi}
\newcommand {\s }   {\sigma}      

\begin{document}
\title     {Global solutions in gravity.}
\author    {M. O. Katanaev
            \thanks{E-mail: katanaev@mi.ras.ru}\\ \\
            \sl Steklov Mathematical Institute,\\
            \sl Gubkin St. 8, 117966, Moscow, Russia}
\date      {}
\maketitle
\begin{abstract}
The method of conformal blocks for construction of global solutions
in gravity for a two-dimensional metric having one Killing vector
field is described.
\end{abstract}
A space-time in gravity models is a differentiable manifold which has
to satisfy at least two requirements. Firstly, the Lorentz signature
metric is to be given on it.
Secondly, the manifold has to be maximally extended along extremals.
The last requirement means that any extremal can be either continued to
infinite value of the canonical parameter in both directions or at a finite
value of the canonical parameter it ends up at a singular point where
one of the geometric invariants, for example, the scalar curvature
becomes infinite. Space-time with a given metric satisfying both
requirements is called a global solution in gravity.
The Kruskal--Szekeres extension \cite{Kruska60,Szeker60} of the
Schwarzschild solution is a well
known example of nontrivial global solution in general relativity.

In general relativity only a small number of global solutions is known
in account of the complicated equations of motion
(see review \cite{Carter73}). In two-dimensional gravity models
attracting much interest last years the situation is simpler, and all
global solutions were found \cite{Katana93A} in
two-dimensional gravity with torsion \cite{KatVol86}, and in a large
class of dilaton gravity models too [6--8].
Constructive method of conformal blocks was proposed \cite{Katana93A}
for two-dimensional gravity with torsion in the conformal gauge.
The use of Eddington--Finkelstein coordinates allowed to prove the
smoothness of global solutions \cite{KloStr96C}. In general relativity
solutions having the form of a warped product of two surfaces can also be
explicitly constructed and classified \cite{KaKlKu99}.
This allowed one to give physical interpretation to many solutions known
before only locally. That is, vacuum solutions to the Einstein equations
with cosmological constant describe black holes, wormholes, cosmic strings,
domain walls, and other solutions of physical interest.

Consider a plane with Cartesian coordinates $\tau,\s$. Let conformally
flat metric of Lorentz signature be given
\begin{equation}                                        \label{emetok}
  ds^2=|N(q)|(d\tau^2-d\s^2),~~~~N\in C^l,~l\ge2.
\end{equation}
Let the argument $q$ to depend on one coordinate only through
an ordinary differential equation
\begin{equation}                                        \label{eshiff}
  \left|dq/d\z\right|=\pm N(q),
\end{equation}
with the following sign rule
\begin{equation}                                        \label{esignr}
\begin{array}{ccl}
N>0: &\z=\s,   &{\rm sign} + {\rm (static)},\\
N<0: &\z=\tau, &{\rm sign} - {\rm (homogeneous)}.
\end{array}
\end{equation}
Conformal factor may have zeroes and singularities in a
finite number of points $q_i$, $i=1,\dots,k$ with $q_1=-\infty$ and
$q_k=\infty$.
We consider power behavior of the conformal factor near $q_i$
\begin{eqnarray}                                        \label{ecfapo}
  |q_i|<\infty:&~~~~&N(q)\sim(q-q_i)^m,
\\                                                      \label{ecfasi}
  |q_i|=\infty:&~~~~&N(q)\sim q^m.
\end{eqnarray}
At $|q_i|<\infty$ for positive $m>0$ the conformal factor equals zero
and defines horizons of a space-time. Negative values $m<0$ correspond
to singularities. At infinite points positive and negative values of $m$
correspond conversely to singularities and zeroes of the conformal factor.
Formulae (\ref{emetok}) and (\ref{eshiff}) define four
different metrics, due to the modulus sign in Eq.(\ref{eshiff}):
\begin{equation}                                     \label{emetdo}
\begin{array}{rl}
  {\rm I}:~~&N>0,~dq/d\s>0, \\
  {\rm II}:~~&N<0,~dq/d\tau<0,\\
  {\rm III}:~~&N>0,~dq/d\s<0, \\
  {\rm IV}:~~&N<0,~dq/d\tau>0.
\end{array}
\end{equation}

The Schwarzschild metric is an example of the type (\ref{emetok}).
Indeed, in static domains {\rm I} and {\rm III} make the coordinate
transformation $\tau,\s$ $\rightarrow$ $\tau,q$
\begin{equation}                                        \label{eschco}
  ds^2=N(q)d\tau^2-\frac{dq^2}{N(q)},~~~~~~N>0.
\end{equation}
Neglecting the angular part in the Schwarzschild metric one gets the
metric (\ref{eschco}) for
\begin{equation}                                        \label{eschcf}
  N(q)=1-\frac{2M}q,
\end{equation}
where $M$ is the mass of the black hole, and $q$
is the radius.

The scalar curvature is the same in all four
domains $R=-N''$. It is singular near $q_i$ for the
following exponents in asymptotic behavior (\ref{ecfapo}):
\begin{eqnarray}                                        \label{esfpsc}
  |q_i|<\infty:&&m<0,~0<m<1,~1<m<2,
\\                                                      \label{esfpcc}
  |q_i|=\infty:&&m>2.
\end{eqnarray}
At $q\rightarrow\pm\infty$ the scalar curvature tends
to a nonzero constant for $m=2$ and to zero for $m<2$.

To construct a global solution we introduce the notion of the conformal
block corresponding to every interval $(q_i,q_{i+1})$. For definiteness
we consider static solution of type {\rm I}.
Then the time coordinate takes values on the whole real axis
$\tau\in R$. The domain of $\s$ is defined by Eq. (\ref{eshiff}) which
may be rewritten as
\begin{equation}                                        \label{eforin}
  \s\sim\int^{q_i,q_{i+1}}\frac{dq}{N(q)}
\end{equation}
This integral converge or diverge depending on the exponent $m$:
\begin{equation}                                        \label{eboucb}
\begin{array}{rl}
  |q_i|<\infty:&\left\{
  \begin{array}{rll}
  m<1,  &{\rm converge,}  &{\rm line,}\\
  m\ge1,&{\rm diverge,}&{\rm angle,}
  \end{array}\right. \\
  |q_i|=\infty:&\left\{
  \begin{array}{rll}
  m\le1,&{\rm diverge,}&{\rm angle,}\\
  m>1,  &{\rm converge,}  &{\rm line.}
  \end{array}\right.
\end{array}
\end{equation}
At the right of this table the form of the boundary of the corresponding
conformal blocks is given. If at both ends of the
interval $(q_i,q_{i+1})$ the integral diverge, then
$\s\in(-\infty,\infty)$, and the metric is defined on the whole
$\tau,\s$ plain. If at one of the ends $q_{i+1}$ or $q_i$ the integral
converge, then the metric is defined on the half plain
$\s\in(-\infty,\s_{i+1})$ or $\s\in(\s_i,\infty)$, correspondingly.
If the integral converge on both sides of the interval, then the
solution is defined on the strip $\s\in(\s_i,\s_{i+1})$.
Next we map the $\tau,\s$ plain on the square along the light like
directions $\x=\tau+\s$, $\eta=\tau-\s$ using a conformal transformation
$u=u(\x)$, $v=v(\eta)$ with bounded functions. Then the static solution
defined on the whole $\tau,\s$ plain corresponds to a square conformal block.
If solution of equation (\ref{eshiff}) is defined on a half interval,
then a static solution corresponds to a triangular conformal block.
When solutions of equation (\ref{eshiff}) is defined on a finite
interval the conformal block is represented in the form of a lens.
There are two conformal blocks for every interval because equation
(\ref{eshiff}) is invariant under the space reflection $\s\leftrightarrow-\s$.

\begin{figure}[htb]
 \begin{center}
 \leavevmode
 \epsfxsize=120mm
 \epsfbox{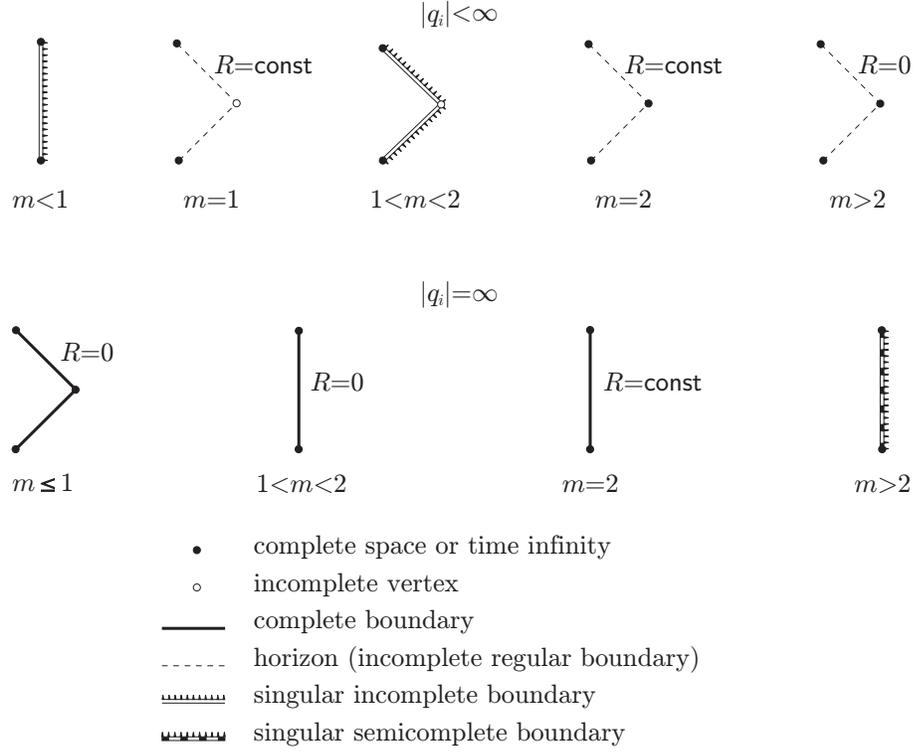}
 \end{center}
 \caption{
 The form of the right boundary of a static conformal block.
 \label{fboupr}}
\end{figure}

A detailed analysis of extremals \cite{Katana00} allows one to
describe all possible boundaries of conformal blocks summarized
in Fig.~\ref{fboupr}.
For definiteness we show the right boundary of static conformal blocks
{\rm I}. Time like boundary is shown by a vertical line
and light like boundary by an angle. If the scalar curvature is singular
on the boundary (\ref{esfpsc}), (\ref{esfpcc}), then it has protuberances.
Incomplete and complete boundaries are shown by dashed and thick solid lines,
respectively. Exclusion is the semicomplete boundary corresponding to
$|q_i|=\infty$ for $m>2$ (see Fig.~\ref{fboupr}). Light like extremals reaching
this boundary are complete while space like ones are not. Lower and upper
corners of all static conformal blocks (time past and future infinities)
are essentially singular points and always complete.
Completeness and incompleteness of right corners of angle boundaries are
shown by filled and not filled circles, respectively.
As a result only horizons correspond to incomplete boundaries, the
scalar curvature being finite on them. Therefore solutions of the form
(\ref{emetok}) must be continued only through horizons.
Let us formulate the rules and the theorem for construction of
global solutions.\\
1) Every global solution for a metric (\ref{emetok}) corresponds to the
interval of the variable $q\in(q_-,q_+)$, where $q_\pm$ are either
infinite points or a curvature singularities defined by the condition
(\ref{esfpsc}). Singularities inside the interval must be absent.\newline
2) If there are no zeroes inside the interval $(q_-,q_+)$ then the
corresponding conformal block is the maximally extended solution.\newline
3) If there are zeroes (horizons) inside the interval $(q_-,q_+)$
then enumerate them, $N(q_j)=0$, $j=1,\dots,n$, $q_j\in(q_-,q_+)$,
and associate with each of the intervals $(q_-,q_1),\dots,(q_n,q_+)$
a pair of static or homogeneous conformal blocks for $N>0$ and $N<0$,
respectively. \newline
4) Sew together conformal blocks along horizons $q_j$, preserving the
smoothness of the conformal factor, that is sew together conformal blocks
corresponding only to adjacent intervals $(q_{j-1},q_j)$ and
$(q_j,q_{j+1})$, and if the gluing is performed for static or
homogeneous conformal blocks, then sew together blocks of one type. \newline
5) The Carter--Penrose diagram obtained by gluing all adjacent
conformal blocks is a connected fundamental region if inside the
interval $(q_-,q_+)$ the conformal factor changes its sign.
If $N\ge0$ or $N\le0$ everywhere inside the interval $(q_-,q_+)$
then one gets two fundamental regions related by space or time
reflection. \newline
6) For one zero of an odd degree the boundary of the fundamental
region consists of boundaries of the conformal blocks corresponding
to the points $q_-$ and $q_+$, and the Carter--Penrose diagram
represents the global solution. \newline
7) If there is one zero of even degree or two or more zeroes of
arbitrary degree the boundary of the fundamental region includes
horizons, and it has to be either continued periodically in space
and (or) time or the opposite sides should be identified. \newline
8) If the fundamental group of the Carter--Penrose diagram is
trivial then it is the universal covering space for a global
solution. \newline
9) If the fundamental group of the Carter--Penrose diagram is
nontrivial, then construct the corresponding universal covering
space. \newline
{\bf Theorem.}
The universal covering space constructed according to the rules 1--9
is the maximally extended pseudo riemannian $C^{l+1}$ manifold with
the continuous $C^l$, metric such that every point not
lying on a horizon has a neighborhood isometric to some domain
with the metric (\ref{emetok}).

The idea of the proof is to go to Eddington--Finkelstein coordinates on
a horizon and to Kruskal--Szekeres coordinates near a saddle point
\cite{Katana00}. All other global solutions are obtained as factor
spaces of the universal covering space by a discrete transformation
group. For example, one may identify similar horizons lying on a boundary
of a fundamental region \cite{Katana93A,KloStr97}.

\begin{figure}[htb]
 \begin{center}
 \leavevmode
 \epsfxsize=120mm
 \epsfbox{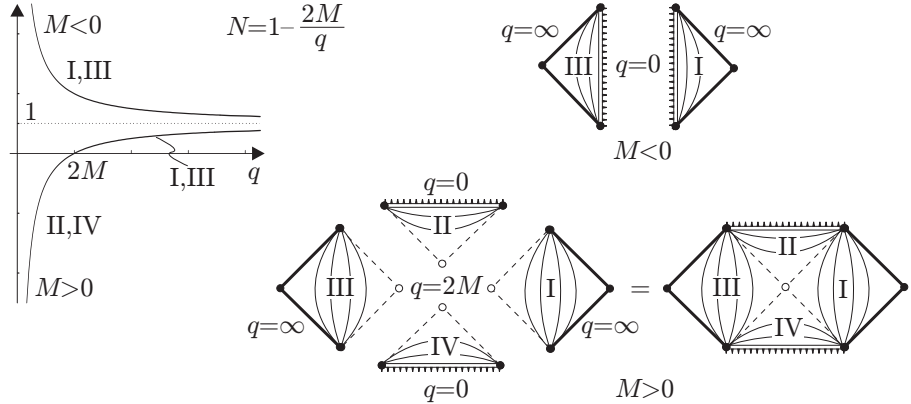}
 \end{center}
 \caption{The conformal factor and Carter--Penrose
 diagrams for the Schwarzschild solution.
 \label{fschws}}
\end{figure}
To illustrate the conformal blocks method we consider the Schwarzschild
solution (\ref{eschcf}) with $M>0$, Fig.~\ref{fschws}. It has a simple
pole at $q=0$ corresponding to a curvature singularity (\ref{esfpsc}).
The point $q_1=2M$ is a simple zero and corresponds to a horizon.
The values $q=\pm\infty$ correspond to asymptotically flat space infinity.
We see that two global solutions for positive and negative $q$
correspond to the infinite interval $q\in(-\infty,\infty)$. For
positive $q\in(0,\infty)$ one has $q_-=0$, $q_1=2M$ and $q_+=\infty$.
For each interval $q\in(0,2M)$ and $q\in(2M,\infty)$ there are two
homogeneous and static conformal blocks.
The boundary elements are defined in Fig.~\ref{fboupr}. The global
solution is uniquely constructed by gluing
together four conformal blocks and yields the Kruskal--Szekeres
extension of the Schwarzschild solution.
For negative $q$ horizons are absent, and maximally extended solutions
are represented by triangular conformal blocks.

The author is very grateful to T.~Kl\"osch, W.~Kummer, T.~Strobl,
I.~V.~Volovich, V.~V.~Zharinov for fruitful discussions. 
This work is supported by Grants RFBR-96-15-96131 and RFBR-99-01-00866

\end{document}